# Simulation of impedance changes with a FEM model of a myelinated nerve fibre

**Ilya Tarotin[1], Kirill Aristovich[1] and David Holder[1]**

[1] Department of Medical Physics, University College London, Gower Street, London WC1E 6BT, United Kingdom

E-mail: ilya.tarotin.14@ucl.ac.uk

## Abstract

*Objective:* Fast neural Electrical Impedance Tomography (EIT) is a method which permits imaging of neuronal activity in nerves by measuring the associated impedance changes (dZ). Due to the small magnitudes of dZ signals, EIT parameters require optimization, which can be done using in silico modelling: apart from predicting the best parameters for imaging, it can also help to validate experimental data and explain the nature of the observed dZ. This has previously been completed for unmyelinated fibres, but an extension to myelinated fibres is required for the development of a full nerve model which could aid imaging neuronal traffic at the fascicular level and optimise neuromodulation of the supplied internal organs to treat various diseases. *Methods:* An active FEM model of a myelinated fibre coupled with external space was developed. A spatial dimension was added to the experimentally validated space-clamped model of a human sensory fibre using the double cable paradigm. Electrical parameters of the model were changed so that nodal and internodal membrane potential as well as propagation velocity agreed with experimental values. dZ was simulated during activity under various conditions and the optimal parameters for imaging were determined. *Main Results:* When using AC, dZ could be recorded only at frequencies above 4 kHz, which is supported by experimental data. Optimal bandwidths for dZ measurement were found to increase with AC frequency. *Conclusion and significance:* The novel fully bi-directionally coupled FEM model of a myelinated fibre was able to optimize EIT for myelinated fibres and explain the biophysical basis of the measured signals.

## 1. Introduction

There is currently increasing interest in exploring ways of imaging fast electrical activity inside peripheral and autonomic nerves. A method that allows visualisation of such neuronal activity will be essential in better understanding normal and pathological states, especially for improving the emerging field of electroceuticals [1], [2] which aims at selective stimulation of nervous tissue and can be applied to peripheral nerves and fascicles that innervate specific internal organs. However, knowing which neuro-anatomical region of the nerve to selectively target is challenging and requires a method for visualisation of fast neural activity such as electrical impedance tomography (EIT). EIT is able to image fast impedance changes during neuronal activity by injecting alternating currents and recording external voltages [3], [4]. This technique has a potential to achieve the goals stated for electroceuticals: it could allow real-time imaging of spontaneous physiological signals to aid selectivity stimulation of the corresponding fascicles and therefore the organs supplied by them for treating various medical conditions. However, even when the whole nerve is activated, the impedance changes measured in EIT recordings are usually very small. Therefore, during less optimal conditions such as physiological spontaneous activity, signal-to-noise ratio (SNR) of the recorded dZ signals is not adequate [4]–[6]. To accomplish successful imaging, the optimal EIT parameters to obtain the largest possible impedance changes must be found. It was previously done for unmyelinated fibres with an accurate FEM model of a C-fibre [7]: the optimized parameters included amplitude and frequency of the injected current, size and location of the electrodes, spatial structure of the fibre and complexity of ion channels. This model was also the first to be able to explain the origin of the positive impedance changes observed at several AC frequencies.

To interpret previously obtained experimental data, optimize EIT for all types of fibres and to have a complete picture in understanding the physiological nature of the observed dZ, a new model of myelinated fibre fully coupled with extracellular space needs to be developed. Together with the model of unmyelinated fibres, this model will be the next building block in the construction of the full model of the nerve consisting of fibres with various geometrical and electrical properties. Besides optimizing EIT, such a model will have a variety of applications like studying the nerve physiology in normal and abnormal conditions and the underlying causes of different illnesses.

*1.1. Fast neural electrical impedance tomography (EIT)*

Fast neural EIT allows imaging neuronal activity during depolarization that occurs with the initiation of the action potential (AP). This technique utilizes injection of alternating currents through two electrodes and recording voltages at the remaining ones. First, numerous voltage measurements are carried out by switching the injection electrode pairs; then, images of the electrical impedance of the tissue are derived from these voltages through solving the corresponding inverse problem using various numerical methods [3]. Fast neural impedance changes are generally 0.01-1% decreases in resistance with the same order of magnitude durations as the action potentials, which are generally shorter in myelinated fibres (~1-2 *ms*) than in unmyelinated (~2-4 *ms*) [8], [9].

The principle of fast neural EIT is based on opening of the ion channels during tissue activation leading to a decrease in its resistance, and this can be measured with the injected AC current. This current starts to pass through open ion channels leading to a change in the recorded voltages proportional to impedance. A decrease of the impedance may be expected to be smaller for higher AC frequencies because of the presence of the capacitance of the lipid neuronal membrane, as. this allows these applied currents to flow through it in both activated and inactivated states of the tissue. This general behaviour was previously predicted with passive [10] and active [7] models of nerve fibres. The latter active model also predicted and explained the finer structure of the dZ decrease with frequency and its increase at certain frequencies as well; it also described the dependence of the dZ on experimental parameters including amplitude and frequency of the injected current and size and position of the electrodes. As a result, active non-linear ion channels interacting with the injected current were shown to be of high importance for precise simulation of the impedance changes, - therefore, a goal to develop an accurate full myelinated model with active ion channels was set in this study.

With EIT, it was possible to visualise neuronal activity in the rat somatosensory cerebral cortex during evoked potentials [11], [12] and ictal epileptiform discharges [13]; prior impedance measurements on unmyelinated crab nerves [5], [14], [15] showed impedance changes up to 1% at DC and an inverse linear relationship to frequency. The recently developed model [7] supported these findings and provided physiological explanations for them.

Fast neural EIT has also been successful in imaging electrically evoked activity of tibial and peroneal fascicles in the sciatic nerve of the rat [4] and the recurrent laryngeal nerve (RLN) of the sheep [16]. The former was achieved by electrical stimulation of the posterior tibial and peroneal nerves and by using a flexible cylindrical microelectrode cuff for transverse AC injection and voltage recording. Measurements were performed at 6-15 kHz with 3 kHz bandwidth; dZ up to 0.2% were recorded and the optimal SNR was found to be at 6 kHz. Imaging of the recurrent laryngeal nerve of the sheep was accomplished in a similar way: following electrical stimulation of RLN, a 28-contact microelectrode cuff was used for AC injection longitudinally along the nerve (9 kHz, 3 kHz bandwidth) and simultaneous activity was recorded in the cervical vagus nerve. Because of the similarity of the experimental designs and the fact that only myelinated fibres were measured in a sciatic nerve and RLN [4], [16], these studies will be used as the main source of data for validation of the myelinated fibre model in this study.

*1.2. Approaches to simulation of myelinated fibres*

Models for simulation of neuronal activity have evolved from voltage-independent resistance [17], [18] to active models, whose resistive properties non-linearly change depending on the transmembrane potential. The newer models include the classic Hodgkin-Huxley (HH) model of the squid giant axon [19] and more recent complex models for unmyelinated [20], [21] and myelinated fibres [22]–[24].

The first active model of a myelinated fibre was based on the classic HH model [25]. As in the HH case, that model contained sodium, potassium and leakage active ion channels that were tuned for the membrane potential to correspond to experimental data obtained in the myelinated nerve fibre of the toad. The model was space-clamped meaning that no propagation of the AP along the fibre was considered. The first spatial simulations of a myelinated fibre were also based on the HH model and included active nodal points with the same ion channels and passive cable internodes with constant resistances [26], [27]. To better replicate experimental data, and since the ion channels utilized in the HH model could predict only basic behaviour of the membrane, new models with more complex ion channels and finer excitability properties started to appear [28], [29] with the addition of ion channels in internodal segments [30], [31].

In addition to ion channels, the spatial structure of the nerve fibre was also revealed to play an important role in the accurate simulation of its excitability properties and shape of nodal and internodal transmembrane potentials. Longitudinal structure was developed from the simple classic cable [28], [29] to the novel paradigm where signals propagated along two cables including axolemma and periaxonal space so that transmembrane and transmyelin potentials could be simultaneously simulated (Fig. 1). Such a double-cable structure was first introduced by Blight [32] and has subsequently been utilized in various studies [6], [22], [31], [33], [34] where it was based on the accurate myelinated fibre morphology obtained using electron microscopy of the paranode-node-paranode region in the cat [35]. In the new models, the incorporation of fine structural features, such as paranodal seal, conducting periaxonal space and multi-layered myelin sheath, were shown to be useful in replication of various excitation properties as depolarizing afterpotentials and the realistic spatial distribution of action potential during conduction [22], [31], [36].

Simulation of fibres together with the external space has been accomplished in several recent studies [37]–[40]. However, these studies involved a consecutive approach where the electrical field in the volume was simulated using FEM, and after that, this electrical field was applied to the compartmental model of the fibre implemented in Matlab or NEURON [41]. This approach is not appropriate when the injected current interacts with the fibre, which in turn simultaneously affects the external space. In this case, a fully bi-directionally coupled FEM model is required.

None of the discussed models allowed estimation of the impedance change associated with the neural activity. The model developed in the current work includes experimentally validated ion channels, simulates realistic behaviour of the human thickly myelinated fibre [23] and enables to continuously measure its impedance during AP propagation.

*1.3. Purpose*

The general purpose of this study was to create an accurate FEM model of a myelinated fibre coupled with extracellular space and predict the dZ during AP propagation under various stimulation paradigms. Specific questions to be answered here were:

1. How does the impedance change depend on the following experimental parameters:
   a. Amplitude and frequency of the injected current;
   b. Signal processing specifications;
   c. size and position of the electrodes.
2. Does this agree with the previous studies?
   a. Does the model validate recent experimental recordings?
   b. Does it offer an explanation on the origin of the dZ?
   c. Does it provide results different from the recent model of unmyelinated fibres [7] and the passive model [10]?
3. What recommendations can be given for optimization of imaging myelinated fibres using fast neural EIT?

The results will improve the understanding of the measured impedance changes and lead to optimization the parameters of EIT for imaging in myelinated nerves, which will facilitate the development of a full model of myelinated and unmyelinated nerve fibres.

## 2. Methods

*2.1. Experimental design*

The work was divided into the following steps:
1. An accurate 1D FEM double-cable model of a mammalian sensory fibre was developed. It contained ten ion channels taken from an experimentally validated space-clamped model [23]: four at the node and six at the internode. Realistic morphology of the fibre [35] was implemented similarly to the one used in [22]. To match the experimental data, several geometrical (Table 1) and electrical (Table 4 in the appendix) parameters were changed in the new FEM model. With these changes, the conduction velocity of the fibre and the shape of membrane potential at the nodes and internodes in resting state and during excitation, matched the experimentally validated space-clamped model [23].
2. The completed 1D model was incorporated into a 3D-equivalent 2D axisymmetric paradigm to form a full coupled model of the fibre with the external space, similarly to the previously developed fully coupled C fibre model (FCCM) [7]. An action potential was induced at the distal end of the fibre and its propagation was simulated intra- and extra-cellularly. External ring electrodes were modelled to apply an electrical current (two electrodes) and to record the axonal activity (one electrode with respect to ground) (Fig. 2). The effects of varying experimental parameters on measured dZ were studied and optimal parameters were established.
3. The dZ simulated in the developed model were compared with the available experimental data on myelinated nerves.

## 2.2. Double cable FEM model of a mammalian thickly myelinated fibre

The simulated myelinated fibre consisted of three types of sections: nodes, paranodes and internodes. This structure partly replicated the one observed by Berthold and Rydmark [35], where authors described nodal (N), myelin sheath attachment (MYSA), fluted (FLUT) and stereotyped internodal (STIN) segments. Compared to this full structure, the FLUT segment of the fibre modelled in this study was combined with the STIN. It is in accordance with [31] where the FLUT region was also omitted and with the MRG model [22] where FLUT region's properties were exactly the same as of the STIN internodal section.

The approach used for development of the model was the finite element method (FEM) which has not been used for simulation of myelinated fibres before. Previously developed compartmental models [22], [31] utilized various forms of the finite-difference approach: the fibre was also divided into smaller parts, but these parts were modelled as passive resistors so that the necessary equations were solved only in the points connecting them; such an approach is used in NEURON software [41]. However, if the purpose is to obtain a solution continuously along the fibre as well as to couple it with an external space in 3D (or, equivalently, 2D axisymmetric) in both directions, the FE M approach is desirable which utilizes nonlinear shape functions approximating the solution of the 2-nd order PDEs at any point in space [42]. The model was built in COMSOL Multiphysics software in conjunction with Matlab, which allowed automatic solving of any partial differential equations using FEM.

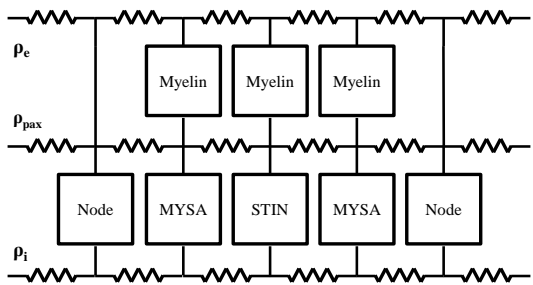

Fig. 1. Circuit diagram of the double cable model of the myelinated fibre developed in this paper and in [22], [31]. Longitudinal resistivities of axoplasm, periaxonal space and external space are designated by $\rho_i$, $\rho_{pax}$ and $\rho_e$. Node, MYSA, STIN and Myelin blocks are represented by parallel capacitance and resistance(s); equations describing the diagram can be derived from Ohm's and Kirchhoff's laws, they are presented below (2); (6)-(9) in the appendix.

A double-cable paradigm was used to simulate the spatial structure of the fibre (Fig. 1). This paradigm showed itself to be able to accurately simulate temporal and spatial distribution of APs in mammalian nerve fibres as well as its various properties such as realistic AP shape, conduction velocity, behaviour in the nodes and internodes [22], [31], [33]. In this model, the axoplasm and the periaxonal space between the axon and myelin sheath were modelled as two parallel cables, which allowed simulation of rapid propagation of APs in the internodes without representing them as perfect insulators.

The geometrical parameters of the modelled fibre were taken from an electron microscopy study [35]; they were very close to the ones used in the MRG model [22]. However, several changes had to be made due to the transition from the previously used compartmental model to FEM (Table 1).

TABLE 1
GEOMETRICAL PARAMETERS OF THE MODEL

| Segment | Parameter | Value |
|---|---|---|
| Node | Length | 1 μm |
|  | Diameter | 5.5 μm |
|  | Mesh | 2 el. |
| MYSA | Length | 3 μm |
|  | Diameter | 12.7 μm |
|  | Mesh | 2 el., $l_2 = 2l_n$* |
|  | Periaxonal space width | 0.02 nm |
|  | N of myelin lamellae | 150 |
|  | Myelin thickness | 2·1.65 μm |
| STIN | Length | 1 μm |
|  | Diameter | 12.7 μm |
|  | Mesh | 10 el., symmetric, exponential** |
|  | Periaxonal space width | 0.2 nm |
|  | N of myelin lamellae | 150 |
|  | Myelin thickness | 2·1.65 μm |
| Fibre | Length | 15 cm |
|  | Full diameter with myelin | 16 μm |
|  | Propagation velocity | 65 m/s |

\* Size of the element adjacent to the node is 2 times smaller
\** Exponential decrease towards the ends of the region, size of the largest central element is 100x size the smallest end elements

Electrical parameters of the model were taken from the realistic space-clamped model of the human sensory fibre [23], but some of them had to be changed for the nodal and internodal APs as well as conduction velocity of the resultant FEM model to correspond to this space-clamped model and experimental results (Table 4 in the appendix). Nodal channels were represented by combined transient and persistent sodium (*Na*) needed for nerve excitation and regulation of subthreshold excitability; fast and slow potassium ($K_{sn}$, $K_{fn}$), Leakage ($Lk_n$) and pump ($P_n$) mainly responsible for membrane potential stabilization. At the internode there were slow and fast potassium ($K_{si}$, $K_{fi}$), hyperpolarization-activated *h-channel*, leakage ($Lk_i$) and pump ($P_i$) whose main functions were to stabilize resting ionic fluxes as well as to regulate repetitive firing and pace-making [43]–[46]. Equations describing the spatial and temporal behaviour of the fibre can be found in the appendix.

To accurately simulate the activity of the fibre with non-uniform structure using the FEM, the mesh must be carefully chosen across elements. For correct simulations, the element size close to the edges of the segments should have been small enough so that the approximating FEM shape functions could accurately simulate transitions of the variables between these segments. Thus, an irregular mesh was used in the STIN regions with the element size exponentially decreasing closer to the points attaching the segments (can be seen in Fig. 2b). The elements' sizes were chosen to be the maximum possible so that the produced solution was constant after they were further decreased; this closely repeated mesh convergence analysis approach [47]. The nodes consisted of 2 equal elements, MYSA – of 2 elements with the element near the node being 2 times smaller than another one, STIN – of 10 elements symmetrically distributed around the centre of the segment so that element length exponentially decreased towards the edges, with the smallest elements being 100 times smaller than the largest element in the centre (Table 1, Fig.2b). Thus, the smallest element in the STIN which was adjacent to MYSA was approximately 2 times larger than the large element in MYSA.

For validation of the developed FEM model, the experimentally validated space-clamped model [23] was reconstructed in MATLAB so that the simulated APs at the node and internode could be matched to the new FEM model via adjustment of its electrical and longitudinal parameters. Also, the conduction velocity (65 m/s) was matched with the experimental datasets that report myelinated sensory fibres of similar size where it was shown to be 50-90 m/s [9], [48]–[50].

To compare the excitability of the model to previously modelled [22], [23] and experimental values [51], the threshold tracking technique called threshold electrotonus was used. The fibre was subjected to 100 ms duration subthreshold depolarizing conditioning stimuli with the amplitude equalling 40% of threshold (defined by 1 ms pulse stimulation); 1 ms test intracellular stimuli were applied to the node each 5 ms in steps ~2% of threshold to trace the threshold during and after the long conditioning pulses.

*2.3. FEM model coupled with external space*

Extracellular space was added to the 1D model to simultaneously simulate AP propagation, injection of the current and recording of the resultant electric field using external electrodes.

The most commonly used approach for simulation of an active fibre and a surrounding space together includes two or three consecutive steps: first, the electric field is simulated using FEM; second, the resultant interpolated voltages are applied to the fibre; third, if necessary, the electric field originating from the fibre is recorded via external electrodes [37]–[40]. This approach was not suitable for our purpose due to complex interaction of injected current with nonlinear ion channels: double-sided coupling was necessary so that the external space can affect the fibre while the fibre can affect the external space. The model utilizing this approach was developed in the authors' previous study on unmyelinated fibres using the FCCM model where its detailed description is presented; the same approach was applied to the myelinated model developed in the current study. As in the case of that study, a 4-electrode impedance measurement paradigm was simulated, which reproduced the setup implemented in a series of experiments of recording impedance changes in unmyelinated crab leg nerves using a linear electrode array [5], [14], [15], [52]. dZ measurements using a silicone microelectrode cuff were accomplished for *in vivo* imaging of myelinated fibres inside rat sciatic nerve [4] and sheep's recurrent laryngeal nerve [16]. AC was applied through electrodes placed transversely across the nerve for rat sciatic and longitudinally for sheep RLN, the same as in the model developed in this study (Fig. 2). In both experiments APs were induced by bipolar stimulation, AC current was injected, and voltages were measured further down the nerve; this basic setup was replicated in the created model.

The external space was modelled by a cylinder (rectangle in 2D axisymmetric space) (Fig. 2a) with the constant electrical conductivity of an extracellular medium equalling 10 *mS/cm* [53]. External ring-shaped electrodes of 0.1 cm width and 0.6 cm diameter were used for current injection and activity recording; the recording electrode was placed 5 cm from the AP initiation point; the injecting ones were at 5.6 and 6.2 cm (Fig. 2a). 2D axisymmetric paradigm, which was shown to provide the same results as the corresponding 3D FCCM model [7], was used to accelerate computations. The fibre was 1-dimensional, which included intracellular, extracellular and periaxonal spaces as well as myelin. AC was injected via two ring external electrodes located on the boundary of the cylinder. An AP was induced at the end of the axon by bipolar stimulation. The nervous activity was recorded by an electrode situated before the injecting ones with respect to ground. The model was grounded at the distal end only so that the current cannot propagate along the fibre in the backward direction which causes artefacts in dZ measurements [14].

The equations for simulation of the external space were similar to the model [7]; they were represented by volume conduction, activating function for modelling the external stimulation and the flux of the membrane current from the fibre to external space. However, their complexity increased due to non-uniformity of the fibre, which required different constants depending on the segment simulated. All the equations are in the appendix, *B*.

A triangular mesh, containing 41,000 elements, was constructed in the axisymmetric model so that the fibre formed a continuous mesh within the volume (Fig. 2b). This means that the lengths of the sides of the triangles adjacent to the fibre were equal to the sizes of the mesh elements in the fibre (Table 1).

In the developed models, equations representing the fibre (1)-(9) (appendix) together with the ones simulating external space and its coupling with the fibre (10)-(13) (appendix) were solved simultaneously for each time step with respect to $V_{ax}$, $V_m$ and $V_e$. For this, an adaptive backward differentiation formula (BDF) was utilized in conjunction with a parallel sparse direct solver (PARDISO) to solve arising linear equations.

Using the voltages recorded during the simulations (Fig. 2), changes in the impedance of the system (*dZ*) were calculated. Details on the simulations setup and signal processing are given in the next subsection and appendix.

### 2.4. Simulation setup and signal processing

In all simulations, current was applied through the injecting electrodes and the voltages were recorded via recording electrode with respect to ground. Each simulation lasted 100 milliseconds and the AP was initiated at t = 50 ms to let the transmembrane and transmyelin potentials stabilize (Fig. 4). The recording sampling rate was 20 kHz when DC and low-frequency AC (225, 625, 1025, 2000 Hz) were applied; it was 100 kHz at all other AC frequencies.

The first step was to study how the impedance of the system changes depending on the amplitude and frequency of the applied currents. For that, the maximum amplitudes of DC and 6 kHz AC current were found at which the dZ measurements stay proportional to them, as expected from the Ohm's law. This revealed the optimal amplitude, which was used in all subsequent simulations; higher currents would modify physiology of the nerve causing artefacts in dZ measurements while too small currents may become comparable to the noise brought by modelling errors or instrumentation [3]. Thus, simulations were performed at 1.3 – 50.2 µA at DC and similarly at 6 kHz in the range of 1.3 – 1256 µA; the optimal amplitudes were chosen.

Using the chosen amplitude, DC and AC currents at a range of frequencies (225 Hz, 625 Hz, 1025 Hz, 2, 4, 6, 8, 10, 12 and 15 kHz) were sequentially applied through the injecting electrodes and the voltages were recorded via the recording electrode with respect to ground (Fig. 2).

The dZ was measured in terms of the recorded voltage:

$$dZ = \frac{Z(t) - Z(t_0)}{Z(t)} \approx \frac{|V(t)| - |V(t_0)|}{|V(t)|} = |dZ| \quad (1)$$

In this equation, the baseline impedance $Z(t)$ changes to $Z(t_0)$ when the AP passes the current injecting electrodes so that dZ represents its relative change. Using the complex form of Ohm's law, the *dZ* and its absolute value *|dZ|* can be written in terms of the measured voltages $V(t)$ and $V(t_0)$. By doing the transition from the impedances to voltages, the phase shift $\Delta\varphi = \varphi_V - \varphi_I$ is supposed to be close to zero as the membrane does not significantly change the phase of the externally measured current [54].

Signal processing for extraction of impedance changes from the recorded voltages included the following (Fig. 3a, Case 1). When AC was applied, then the two sequential simulations in phase (0, π/4, π/2, 7π/8, 5π/8, 19π/20) and in antiphase (-π, -3π/4, -π/2, -π/8, -3π/8, -π/20) locked to the time of AP initiation were carried out at each frequency (Fig. 3). This was done for validation of the obtained dZ and removal of the artefacts appearing due to coherence of AC with the AP. The resultant signals were 1) subtracted, which led to cancellation of the AP and doubling of the dZ; 2) band-pass filtered around the carrier frequency, to eliminate the AP artefacts left after subtraction; 3) demodulated, using the absolute of Hilbert transform and 4) normalized in respect to baseline, to obtain normalised dZ in percent. In the simpler DC case, the signals simulated with the reversed polarity of the current were subtracted and normalised.

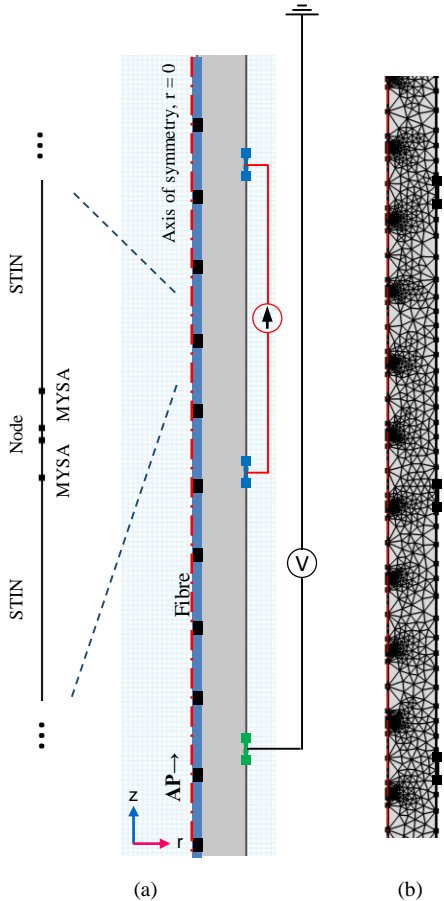

Fig. 2. (a) The 2D axisymmetric geometry of the model.
The fibre is depicted by a blue line with black nodal and MYSA regions, the axis of symmetry is shown by the red dash-dotted line. The AP was induced from the end of the fibre; DC or AC were applied through two external electrodes (blue) placed 5.6 cm and 6.2 cm from the axon's end; the electric field was recorded by an external electrode (green) placed before the injecting ones, 5 cm from the proximal end of the fibre (Table 2). A magnified image of the fibre in the region around the node is shown on the left; its correspondence to the 2D axisymmetric model is depicted by the blue dashed lines;
(b) Triangular FEM mesh of the model combined with the nonregular mesh of the fibre. Elements' sizes are provided in *Methods, B* and *Table 1*.

Impedance change measurements were also performed when subtraction (*step 1*) of the signal processing was omitted (Fig. 3b, case 2): phase and antiphase recordings were not subtracted, the original signal containing extracellular AP was band-pass filtered and demodulated (Fig. 3, Case 2). This was done in order to check the feasibility of real-time imaging in experimental conditions: in the case 2 only one recording was necessary to obtain a dZ while subtraction demanded in- and anti-phase measurements to be performed.

In contrast to the previous modelling and experimental studies [4], [7], [16] where constant bandwidths were used across all injected AC frequencies, they were varied in this study to find the optimal one at each frequency. Since the durations of simulated APs were short (< 1 *ms*) (Fig. 3, Fig. 4) and in order to have sufficient time resolution of the obtained dZ signal – high bandwidth were better to be used during the band-pass filtering step of signal processing.

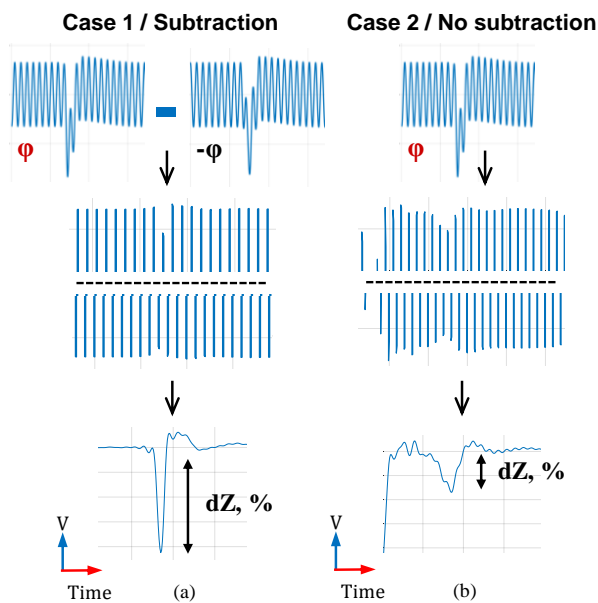

Fig. 3. Signal processing to extract the dZ. It was performed for two cases: (a) with and (b) without subtraction of signals simulated with the in-phase and anti-phase injected AC (presented at the top of the figure). After this step, the signal was band-pass filtered (top and bottom parts of the resultant sine wave are in the middle of the figure), demodulated and normalized (at the bottom of the figure).

In fact, dZ measured at DC possesses the full natural frequency band, which reflects the real apparent dZ; it could be reached at a given AC frequency if time resolution is sufficient and a full dZ bandwidth is taken into account when processing the signal at this frequency. On the other hand, it is not always possible to choose bandwidths covering the characteristic frequency of the real dZ (0–8 kHz, Fig. 7). The first reason for this is that the carrier frequency has to be high in order to support this high bandwidth. Second, if recordings with in- and anti-phase currents were not subtracted and the frequency of the carrier is sufficient, necessary bandwidth would start to overlap with the characteristic frequency band of the AP (1–2 kHz, Fig. 7), which brings artefacts into measurements (Fig. 9a in Results). Finally, as the membrane consists of parallel resistances and capacitance, the dZ will decrease with the frequency of the measuring current. All of these factors contribute to the measured dZ, so that optimal values of AC carrier frequency and bandwidth must be found

At each frequency higher than 625 Hz, dZ was obtained at a range of bandwidths from 500 Hz up to ($f$ – *500*) Hz with 100 Hz steps, where *f* is the frequency of the applied AC. This was repeated for each phase-antiphase pair and the optimal bandwidth was determined as the one that maximizes the value $\Delta = (dZ - 3 \ standard \ deviations \ of \ dZ)$. Three standard deviations of dZ (obtained by measuring at different phases of injected AC) were chosen because: 1) the magnitude of the obtained dZ increases with bandwidth due to presence of the high frequency components in it; 2) the standard deviation of dZ also increases because the AP frequency band (0–2 kHz, as seen from the power spectral density of the modelled AP (Fig. 7)), starts to overlap with the chosen bandwidth; in this case, the dZ measurement become unreliable. The above procedure was repeated in situations when subtraction of in-phase and anti-phase signals (*step 1* of the signal processing) was omitted (Fig. 3b), which resulted in the errors becoming much larger: even very small overlap in the bandwidth and the AP frequency band brings huge errors to measurements. Therefore, this overlap should be minimized by decreasing the utilized bandwidths that then leads a reduction in the recorded dZ.

Further, the locations and sizes of the electrodes were varied to study their effect on the recorded dZ, as has previously been done for the unmyelinated FCCM model. The varied parameters included the diameter of the electrodes (and subsequently surrounding volume) $D_{el}$, width of the electrodes $H_{el}$, distance between recording and injecting electrodes $\Delta x_R$ and distance between injecting electrodes $\Delta x_I$ (Table 2). Variation was done at DC as well as 6 kHz which was chosen due to the dataset on the sheep recurrent laryngeal [16], with the optimal bandwidth found during the study (Fig. 8).

TABLE 2
GEOMETRICAL PARAMETERS OF THE COUPLED MODEL

| Parameter | Value |
|---|---|
| Diameter of the electrodes / surrounding volume, $D_{el}$ | 0.2 cm*; 0.2 – 2 cm |
| Width of the electrodes, $H_{el}$ | 0.02 cm; 0.01 – 0.1 cm |
| Distance between recording and injecting electrodes, $\Delta x_R$ | 0.6 cm; 0.1 – 5 cm |
| Distance between injecting electrodes, $\Delta x_I$ | 0.6 cm; 0.1 – 5 cm |

*First number in each row is the value used by default; range of values shows a variation of the parameter (see Fig. 10 in the results).

Finally, to understand the origin of the measured dZ, membrane conductances, the flow of applied currents through the membrane and the extracellular space were studied. The gradient field of the current was plotted at DC so that its direction could be seen in any point of the external space; it was done in two time points – when negative and positive dZ reached maximum (50.7 and 51.5 ms, Fig. 6a, Fig. 9a). Conductances and membrane currents were measured along 2 cm of the fibre under the electrodes at 4.8 – 6.8 cm from the point of AP initiation (Fig. 2).

## 3. Results

### 3.1. Double-cable FEM model of a single myelinated human sensory fibre

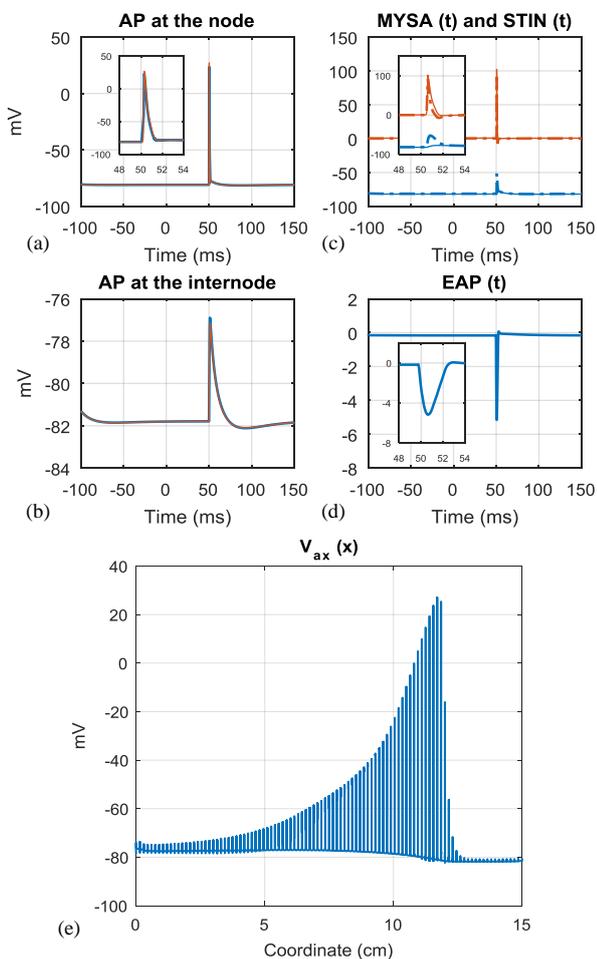

Fig. 4. Action potentials in time (a-d) and along the length (e) simulated with the developed model and compared to the validated space clamped model. 250 ms long simulations are presented to show variations in resting potential preceding and following the AP. Magnified plots of 6 ms durations are embedded into (a), (c), (d) to better reveal the shapes of the depicted signals.
(a) Transmembrane potentials measured at the centre of the nodal region simulated with the developed FEM model (red lines) in comparison with the validated space-clamped model [23] (blue lines).
(b) Transaxonal potential measured at the centre of internodal region (red) compared with the validated model (blue);
(c) Transaxonal (blue) and transmyelin (red) potentials at the MYSA (dashed lines) and STIN (solid lines) regions;
(d) AP recorded with the external electrode in respect to ground, EAP (t);
(e) Transaxonal potential along the fibre length, Vax (x).

The developed one-dimensional FEM model was validated against the existing experimentally validated space-clamped model [23] so that the shape and amplitude of APs at the node and internode in these models match each other (Fig. 4). Conduction velocity in the model equalled to 65 *m/s* which is in the experimental range for the myelinated fibre of similar properties [9], [48]–[50].

The duration of the AP at the node was equal to approximately 1 ms, while depolarization at the internode lasted around 20 ms; the amplitudes at the node and internode were ~110 mV and 5 mV respectively (Fig. 4).

APs at various segments of the fibre (Fig. 4) were similar to the results obtained in the previous modelling studies [22], [31], [37]. Nodal and extracellular APs (Fig. 4a, c) were also in agreement with the experiments [30], [55]; no experimental data is available for transmyelin potentials and internodal values of transaxonal potentials. The spatial length of the AP was ~12 cm (Fig. 4e) which was similar to the values obtained in [31], [36].

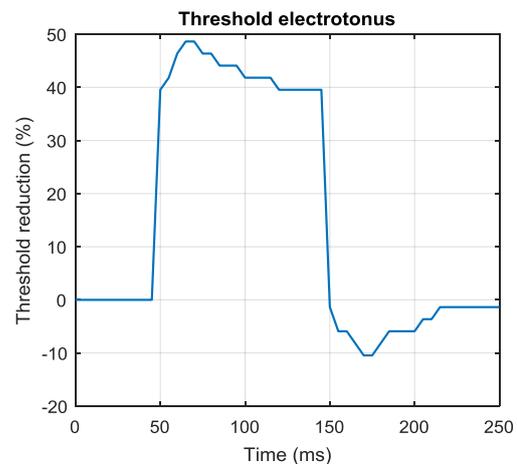

Fig. 5. Fibre excitability – threshold electrotonus. It was created in two steps: 1) a long 100-ms conditioning stimulus with the 40% of threshold amplitude was applied; 2) threshold for AP initiation was found by application of 1-ms long stimuli to the node every 5 ms during and after the conditioning stimulus. The graph is not smooth because short 1-ms threshold determining stimuli were applied in steps equalling approx. 2% of threshold (~25 pA).

The model's threshold electrotonus was similar to the ones obtained in recent modelling and experimental studies [22], [23], [51]. It included transient threshold changes following the start (40 to 49% threshold reduction at 50-70 ms) and the end of the stimulus (-1 to -10% reduction at 150-170 ms) (Fig. 5).

## 3.2. Full model of a myelinated fibre bi-directionally coupled with external space

### 3.2.1. Optimal current amplitude

The simulated dZ were linear with the injected direct currents from 1.3 – 12.6 µA: the shape and percentage dZ stayed the same in this range (Fig. 6a). dZ became nonlinear at higher currents – these currents change normal behaviour of the membrane; for example, it can be seen at 50.2 µA (Fig. 6a). The upper safe range limit increased at higher frequencies: for example, at 6 kHz dZ linearity remained up to 125 µA (Fig. 6b). Thus, the membrane was not affected by currents of up to 12.6 µA at all frequencies, therefore this amplitude was chosen for all further simulations.

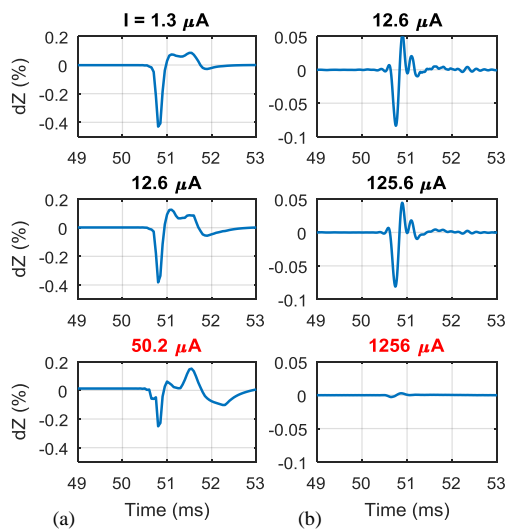

Fig. 6. Impedance changes at different current amplitudes measured at (a) DC, left column; (b) 6 kHz, right column.
Currents at which dZ become nonlinear are shown with titles highlighted in red. Time markers during simulation: AP excitation – 50 ms from the start; AP passes under the recording electrode – 50.8 ms; AP reaches the end of the fibres in 2.3 ms after start.

### 3.2.2. Optimal current bandwidth and frequency

The power spectral density of dZ consisted of two main components – 1 kHz (corresponding to positive dZ) and 2 kHz (negative dZ) and included the band of up to 8 kHz (Fig. 7). Therefore, the filtering bandwidth used during signal processing for subsequent demodulation (Methods, *D*) could not be significantly less than 2 kHz that restricted the lower limit of the carrier frequencies.

When signal processing involved the subtraction of in-phase and anti-phase recordings, the optimal bandwidths to obtain the largest signal were found to increase with the carrier frequency from around 500 Hz at 4 kHz AC to 4.5 kHz at 6 kHz AC, 7.5 kHz at 8 kHz AC, 8.1 kHz at 10 kHz AC, 9 kHz at 12 kHz AC and 10.6 kHz kHz at 15 kHz AC (Fig. 8, Table 3).

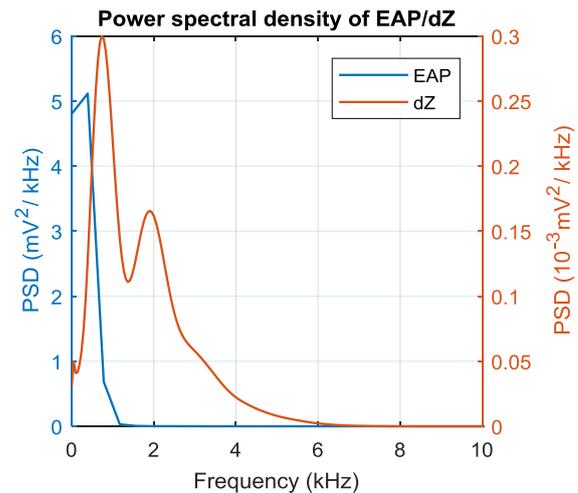

Fig. 7. Power spectral density estimate of the simulated action potential (Fig. 4a) and dZ at DC (Fig. 9a). PSD of dZ simulated at DC demonstrates the frequency band of a natural dZ; it could be reached at AC if a perfect carrier frequency and bandwidth are chosen (see text).

When no AP subtraction has been performed, significant dZ could only be recorded above 6 kHz (Fig. 8, Fig. 9). The optimal bandwidth increased from 0.9 kHz at 6 kHz AC to 1.1 kHz at 8 kHz AC, 2.1 kHz at 10 kHz AC and 3.9 kHz at 8 12 kHz AC; but it decreased to 1.8 kHz at 15 kHz AC (Fig. 8, Table 3).

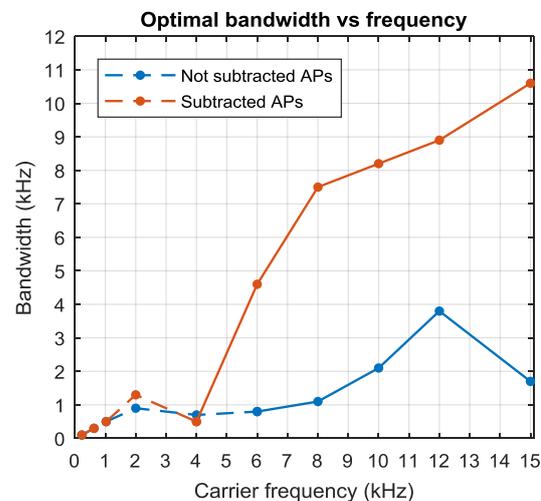

Fig. 8. Optimal bandwidth providing the highest reliable dZ signal for each injected AC frequency. It was calculated when the value *(dZ – 3 standard deviations)* reached a maximum. The case when phase-antiphase subtraction was performed as a 1st step of signal processing is shown in red, the alternative case is depicted in blue. Dashed lines show that no reliable impedance changes could be obtained at < 4 kHz for the subtraction case or < 6 kHz for the case when no subtraction was accomplished.

For the initially chosen geometrical parameters (Table 1, Table 2), significant negative dZs simulated at optimal bandwidths when subtraction was performed were ($\cdot 10^{-2}$): -45±4% at DC, -0.7±0.3%, -9.6±1.5%, -11.4±0.9%, -

<mark>Journal XX (XXXX) XXXXXX</mark> <mark>Author *et al*</mark>

8.0±0.9%, -6.3±0.4%, -5.4±0.6% at 4, 6, 8, 10, 12 and 15 kHz (Table 3). Significant dZ increases were also observed (·$10^{-2}$): 6±4% at DC and 4.2±1.5%, 4.5±0.9%, 1.8±0.9%, 1.3±0.4%, 1.3±0.6% at 6, 8, 10, 12 and 15 kHz (Fig. 9b). The analysis method without subtraction was not available at DC as this requires different polarities of current. When subtraction was not performed, significant dZ decreases were (·$10^{-2}$): -1.9±0.9%, -1.9±0.4%, -3.1±0.6%, -4.4±0.6%, -2.3±0.4% at 6, 8, 10, 12 and 15 kHz; dZ increases were observed to be significantly different from zero at 10, 12 and 15 kHz with the values of 0.8±0.6%, 1.4±0.6%, 0.6±0.4% ·$10^{-2}$ respectively (Fig. 9c, Table 3).

TABLE 3
MAIN SIMULATION RESULTS

| Signal processing | Carrier frequency | Optimal bandwidth | dZ* decrease (mean ± s.d.) | dZ* increase (mean ± s.d.) |
|---|---|---|---|---|
| In- / anti-phase subtraction | DC | - | -0.45±0.04% | 0.06±0.04% |
| | 4 kHz | 500 Hz | -(0.7±0.3)·$10^{-2}$% | - |
| | 6 kHz | 4.5 kHz | -(9.6±1.5)·$10^{-2}$% | (4.2±1.5)·$10^{-2}$% |
| | 8 kHz | 7.5 kHz | -(11.4±0.9)·$10^{-2}$% | (4.5±0.9)·$10^{-2}$% |
| | 10 kHz | 8.1 kHz | -(8.0±0.9)·$10^{-2}$% | (1.8±0.9)·$10^{-2}$% |
| | 12 kHz | 9 kHz | -(6.3±0.4)·$10^{-2}$% | (1.3±0.4)·$10^{-2}$% |
| | 15 kHz | 10.6 kHz | -(5.4±0.6)·$10^{-2}$% | (1.3±0.6)·$10^{-2}$% |
| No subtraction ("single shot") | DC | - | - | - |
| | 4 kHz | - | - | - |
| | 6 kHz | 0.9 kHz | -(1.9±0.9)·$10^{-2}$% | - |
| | 8 kHz | 1.1 kHz | -(1.9±0.4)·$10^{-2}$% | - |
| | 10 kHz | 2.1 kHz | -(3.1±0.6)·$10^{-2}$% | (0.8±0.6)·$10^{-2}$% |
| | 12 kHz | 3.9 kHz | -(4.4±0.6)·$10^{-2}$% | (1.4±0.6)·$10^{-2}$% |
| | 15 kHz | 1.8 kHz | -(2.3±0.4)·$10^{-2}$% | (0.6±0.4)·$10^{-2}$% |

*Only dZ significantly different from zero are shown

### 3.2.3. Influence of the size and position of the electrodes

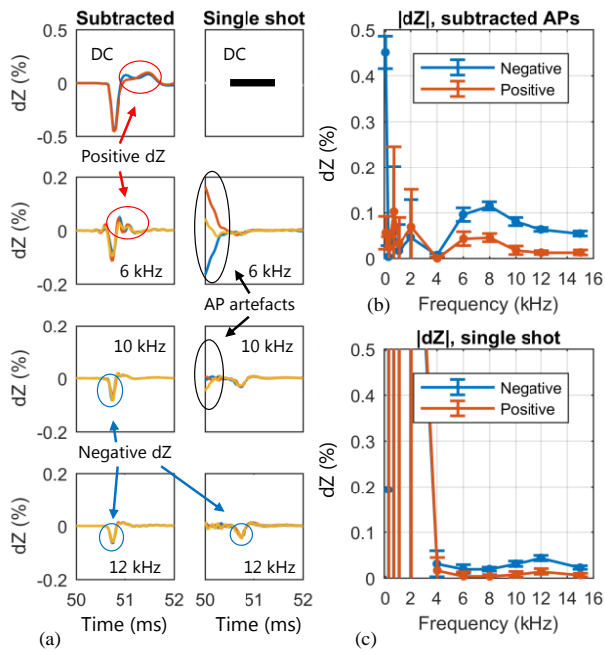

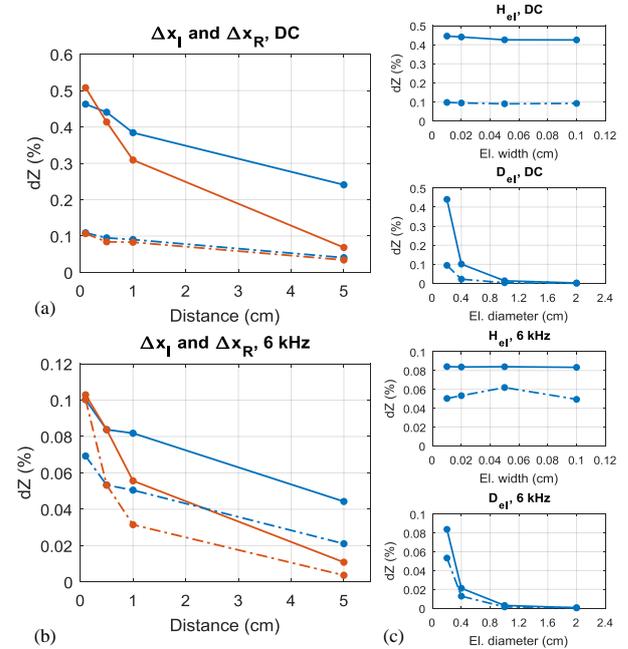

Fig. 9. (a) Examples of impedance changes (dZ) at DC, 6, 10 and 12 kHz for subtraction and single shot cases (Fig. 3). It was not possible to measure dZ at DC without AP subtraction with the used signal processing (Fig. 3). Lines of different colours represent dZ recorded at different phases or different polarity (for DC) of the injected current. Examples of negative dZ are highlighted by blues circles, positive ones – by red circles; the artefacts arising from APs in a single shot case are highlighted by black circles;
(b) Absolute dZ vs. frequency at optimal bandwidths (Fig. 8) for the case when in- and anti-phase AP were subtracted during signal processing;
(c) Absolute dZ vs. frequency at optimal bandwidths when no subtraction was performed. Blue lines designate impedance decrease (negative change), red – impedance increase (positive change). Error bars are standard deviations calculated for the dZ simulated at different phases of the current (AC) and at different polarities (DC). Instability in (c) below 4 kHz is observed as it is impossible to demodulate dZ from the carrier frequency significantly lower than the characteristic frequency band of the dZ itself (Fig. 7).

Fig. 10. Absolute impedance changes versus size and position of electrodes. dZs were obtained subtraction of APs (Fig. 3, case 1).
Dependence on distance between the recording and injection electrodes ($\Delta x_R$) and between injecting electrodes ($\Delta x_I$) at (a) DC and (b) 6 kHz with bandwidth = 4.6 kHz (optimal for 6 kHz, Fig. 8).
dZ vs $\Delta x_R$ is depicted with blue lines, $\Delta x_I$ – with red lines. Negative dZ are shown by full lines, positive – by dashed lines.
(c) Dependence of dZ on width ($H_{el}$) and diameter ($D_{el}$) of the electrodes. Top two graphs are for DC, bottom graphs – for 6 kHz AC. Negative dZs are shown by solid lines, positive – by dashed lines.

At both DC and 6 kHz, which was chosen due to the dataset on the sheep recurrent laryngeal, when distances between injecting ($\Delta x_I$) and the recording and injecting electrodes ($\Delta x_R$) were increased, the positive and negative dZ decreased (Fig. 10a, b).

At DC, the maximum negative dZ of -0.47% and -0.5% was simulated at $\Delta x_R = 0.1$ cm and $\Delta x_I = 0.1$ cm; the maximum positive dZ was about 0.1% at $\Delta x_I$ or $\Delta x_R = 0.1$ cm. At 6 kHz, the negative maximum values were -0.1% at 0.1 cm $\Delta x_I$ or $\Delta x_R$, the positive ones were 0.07% and 0.1% at the same electrodes' locations. The absolute dZ values significantly decreased with increasing diameters ($D_{el}$) of the electrodes up to 2.0 cm but stayed close to constant when increasing their widths ($H_{el}$) up to 0.1 cm (Fig. 10c).

### 3.2.4. Biophysical origin of the recorded dZ

During the propagation of AP, the overall membrane conductance increased significantly (Fig. 11a): the main contribution was made at the node where it increased from 0.025 to 0.57 S/cm$^2$ (~23 times). The total increase was approximately the same: compared to the nodal values, the contribution of the internodal axolemma and myelin was negligible. This conductance increase could explain the dZ decrease which was being measured externally and occurred at the same time (Fig. 6, Fig. 9a).

The total change in conductance induced by the injected EIT direct current (Fig. 11b) was about -2.1 *mS/cm$^2$*, which was less than 0.5% of the overall change in conductance during an AP. Therefore, the injected current did not significantly interfere with the membrane and affect the dZ measurements.

As the largest conductance change during AP was at the nodal regions (Fig. 11a), the flow of the nodal currents (Fig. 11c), as well as the injected EIT current through nodes (Fig. 11d), were studied. The main contributors to the nodal currents were the ionic currents whose absolute values rose from ~0 to 18 mA/cm$^2$ (Fig. 11c). The flow of the applied DC through nodes (Fig. 11d) increased synchronously with dZ decrease (Fig. 6, Fig. 9a); it was followed by a decrease in the flow in respect to baseline seen during dZ increase (Fig. 11d). Thus, less current passed through the nodal membrane during an impedance increase when compared to the resting state.

The gradient field of the current constructed at the time when the dZ decrease and increase reached maximum (50.7 and 51.5 ms from the start of simulation, the AP was launched at t = 50 ms, *Methods, D*) showed that the direction of the current flow at the position were recording electrode was located (5 cm), was different during the dZ decrease and during its increase (Fig. 11e, f). Voltage measured by the recording electrode (in respect to ground) was modulated by the impedance only if the current was constant (V=Z/I, eq. (1)); however, the current redistributed in the external space during different phases of AP propagation (Fig. 11e and Fig. 11f) thus affecting the dZ.

The injected EIT current flow through the nodes (Fig. 11d) matched estimates from the gradient field (Fig. 11e, f): the direction of current flow changed at the area where the recording electrode was located (green on Fig. 11e and f) and it was accompanied by a decrease in the current flow through this area. Thus, the positive dZ was not a natural increase in membrane impedance but was observed due to redistribution of the applied current flow resulting from the membrane activity.

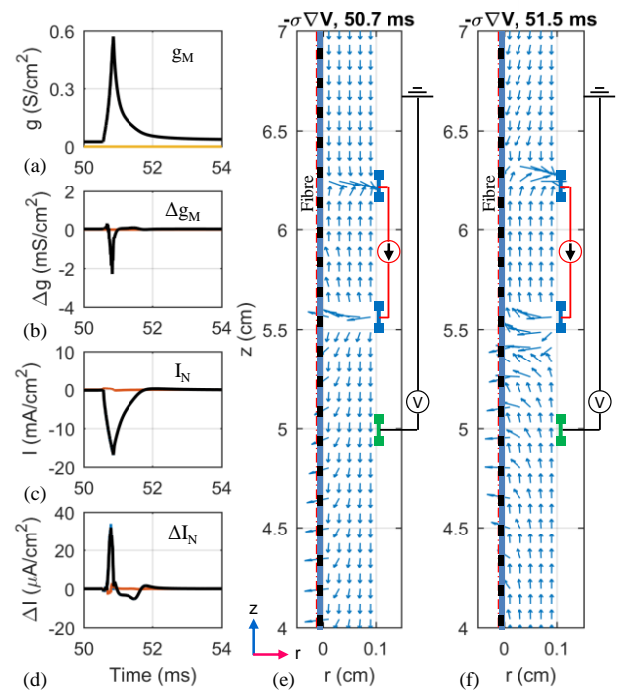

Fig. 11. (a) Conductance of the fibre during an action potential when DC was applied. Nodal values are depicted by the blue line, internodal (MYSA + STIN) – by red, myelin – by yellow and total values – by black thick lines;
(b) Change in conductance induced by the applied DC, $\Delta g_M = g_{DC}$-$g_0$; colour legend is the same as in (c);
(c) Membrane current flow at the node measured at DC. Blue line denotes the sum of ionic currents, red line – capacitive current, black line – total current;
(d) Change in membrane current flow induced by the applied DC through the node, $\Delta I_N = I_{DC}$-$I_0$; colour legend is the same as in (f);
(e) Distribution of the current flow (**j** = -σ∇V) at the time when the negative component of dZ was maximal (50.7 ms, Fig. 6a, Fig. 9a). Arrows designate the direction of the flux at each point of the extracellular volume (Fig. 2). The gradient current field is normalized, i.e. all the arrows have the same length;
(f) Distribution of the current flow (j = -σ∇V) at the time when the positive component of dZ is maximal (51.5 ms, Fig. 6a, Fig. 9a).

### 3.3. Comparison with experimental data

The negative dZs simulated using high-frequency AC currents and their dependence on frequency were in close agreement with data recorded from the sciatic nerve of the rat (Fig. 9b, c) [4]. In that work, dZ SNR was studied at 6, 8, 10, 11 and 15 kHz; also, the phase of the applied AC current was randomized, so it could be compared to the case when in phase and anti-phase recordings were subtracted (Fig. 9a). The maximum SNR = 8 was found to be at 6 kHz, it decreased to 2.2 at 8 kHz, but increased again to 4 at 10 kHz, 7.1 at 11 kHz and 5.4 at 15 kHz. Taking into consideration the root-mean-square experimental noise which was constant at 6, 11 and 15 kHz (0.5±0.01 µV) and at 8 and 10 kHz (2.5±0.1 µV) [4], it is seen that the pure signal was the largest at 10 kHz. In spite of general agreement, the differences between the experimental and simulated results can be explained by: 1) the whole nerve was used instead of a single fibre which changes the compound AP latency due to dispersion in nerve and 2) in the experimental results, a bandwidth of 3 kHz was used for all frequencies, which this current study has revealed is not the optimal parameter setting to use (Fig. 8). Additionally, positive changes have not been recorded experimentally in myelinated nerves likely because of their small magnitudes which makes it difficult to obtain adequate SNR due to the noise and dispersion absent in single-fibre simulations [4], [56].

A number of studies of dZ measurements in unmyelinated nerves of the crab exist, testing from DC [15], [52], [57] and up to 1025 Hz AC [5], [14]. In those studies, the impedance changes were the largest at DC and decreased with AC frequency, which partly agrees with the simulations (Fig. 9) – while the dZ was the largest at DC, it was impossible to measure it at AC frequencies lower than 4 kHz because of the high characteristic frequency of the AP of the modelled myelinated fibre (Fig. 7). The dZ did not decrease with frequency from 4 to 15 kHz as the optimal bandwidths for each frequency were used instead of constant bandwidth utilized in all the studies listed above.

The distances between electrodes have been experimentally studied in unmyelinated nerves only; they agree with the simulated results for the distance between injecting electrodes ($\Delta x_I$) [5], [15], [57] but differ from the ones for the distance between recording and injecting electrodes ($\Delta x_R$) [15], [52] (Fig. 10). The dZ dependence on the size and locations of the electrodes obtained in this study is in a full agreement with the ones simulated with the C fibre FCCM model [7].

The threshold electrotonus of the developed model (Fig. 5) closely corresponds to the one obtained in the caudal nerve of the rat [51] as well as the recent validated models of myelinated fibres [22], [23].

## 4. Discussion

### 4.1. Summary of results

1. The developed 1D FEM model of a myelinated sensory fibre was validated against an experimentally validated space-clamped model [23] and agreed with experimental data on AP shape at the node and internode as well as on propagation velocity and threshold electrotonus [9], [48]–[51].

2. At all frequencies, the simulated dZ did not depend on the amplitude of the applied current if it was less than 12.6 µA.
3. The dZ simulated with the developed bi-directionally coupled FEM model of a myelinated fibre were maximal at DC and were different if the in-phase and anti-phase signal were subtracted during signal processing or not (Fig. 3). In both cases, dZ could not be simulated at frequencies lower than 4 kHz due to the high characteristic frequency of the AP and dZ (Fig. 7). When subtraction of in- and antiphase signals was performed, optimal bandwidths for obtaining the largest dZ were increasing with frequency from 500 Hz at 4 kHz AC to 10.6 kHz at 15 kHz AC (Fig. 8); magnitudes of dZ decrease reached maximum value equalling -0.11% at 8 kHz (Fig. 9b). In case when the signal was band-pass filtered without preliminary subtraction, the optimal bandwidth increased from 0.9 to 3.9 kHz at 6 to 12 kHz AC and decreased to 1.8 kHz at 15 kHz AC (Fig. 8); the maximum dZ decrease was -0.04% at 12 kHz in this case (Fig. 9b). Small dZ increases resembling the same behaviour with frequency were also recorded (Fig. 9). These simulations partly agreed with recent modelling [7], [22], [23] and experimental results [4], [5], [15], [52], [57].
4. The simulated dZ decreased with the distance between the injecting electrodes ($\Delta x_I$) and between recording and injecting electrodes ($\Delta x_R$) (Fig. 10a, b). For $\Delta x_I$, it was in accordance with experiments carried out on unmyelinated fibres [5], [15], [57], but the dependence was different for $\Delta x_R$. The dZ decreased significantly with diameters of the used electrodes and stayed constant with their widths (Fig. 10c). All these results fully agreed with the unmyelinated fibres FCCM model.
5. The origin of the negative impedance change recorded during neural activity was the significant increase in the membrane conductance associated with this activity (Fig. 11a). The impedance increase following the decrease at all frequencies (Fig. 6, Fig. 9a) appeared due to redistribution of the injected current at the area of recording (Fig. 11e, f). In this case, the recorded voltage reflected not only the change in impedance but also a change in the direction of the current flow induced by the membrane activity.

### 4.2. Answers to the stated questions

1) How does the impedance change depend on the experimental parameters?

   *a. Amplitude and frequency of the injected current;*

   The maximum current at which dZ was linear with it was 12.6 µA when DC was applied; the limit significantly increased at AC equalling approximately 125 µA at 6 kHz (Fig. 6). The linearity of the dZ with the injected current is important for the measured voltage to signify impedance change and not depend on the current itself. As a result, 12.6 µA was chosen as a safe current which did not affect neuronal excitability and was used in all simulations.

The largest dZ was simulated at DC; at AC, the optimal frequency providing the largest dZ depended on the way of signal processing. In the case when subtraction of in-phase and anti-phase signal was performed (Fig. 3), the largest dZ was observed at 8 kHz, in the other case, the optimal frequency was 12 kHz but the dZ was ~3 times smaller. These values were obtained using the optimal bandwidths for each frequency (Fig. 9). They were chosen as the ones providing the largest value of ($dZ – 3·S.D.$) when various phases of the current locked to the AP initiation were applied (*Methods, D*).

### b. Signal processing specifications

Two signal processing scenarios was studied: 1) when two signals with the current in-phase and in antiphase were recorded and then subtracted before band-pass filtering and demodulation; 2) "single-shot" measurement, when the raw recorded signal was band-pass filtered around the carrier frequency and demodulated (Fig. 3). The first scenario resulted in higher dZ because the AP was subtracted and therefore the artefacts associated with it (Fig. 9) were absent. For this case, the optimal parameters were 8 kHz AC with 7.5 kHz bandwidth (Fig. 8, Fig. 9): there was no AP artefact and the large bandwidth allowed to extract full frequency band of the dZ (Fig. 7). The second case when subtraction was not performed was studied because it could provide the means for real-time imaging as there was no need to record twice, in phase and in antiphase, to obtain a dZ. In this case, the possible bandwidth was limited because the AP was present (Fig. 7) causing artefacts in the measurements. Thus, the optimal parameters were to use 3.9 kHz bandwidth at 12 kHz AC (Fig. 8, Fig. 9).

The reason for the increase in the optimal bandwidth with the measuring AC frequency is that the dZ has frequency components at up to 7-8 kHz that are resolved with higher temporal resolution and therefore, bandwidth. The optimal bandwidth fell from 12 to 15 kHz AC when there was no subtraction (Fig. 8) because the natural dZ of the fibre decreased with frequency (since the membrane is a parallel resistance and capacitance) while standard deviation increased with it. This increase was due to high-frequency noise brought by modelling errors; this noise was the same in different simulations and was eliminated during phase-antiphase subtraction. Although this noise is absent in experiments, the "single shot" model case could provide a general trend showing an increase in the optimal bandwidths' values with the carrier frequency up to 12 kHz (Fig. 8), which is important for optimizing signal processing for fast neural EIT.

### c. Size and position of the electrodes

With increasing distance between the electrodes, the decrease in dZs was observed (Fig. 10a, b); dZ decreased significantly with increasing electrodes diameter and was not affected by their widths (Fig. 10c). These results were in agreement with the ones for unmyelinated fibres FCCM model.

2) Does this agree with the previous studies?

### a. Does the model validate recent experimental recordings?

The modelled frequency dependence of the dZ decrease is in general agreement with the experimental data on the sciatic nerve of the rat [4]. Although the model was developed for a single fibre, the results were comparable as dZs were shown to have a close-to-linear behaviour with the increasing number of fibres [7].

Also, as real nerves consist of thousands of fibres, there is an effect of temporal dispersion [56] – increase in the latency of the compound AP which occurs due to variability of fibres' conduction velocities. This affects the dZ latency in the same way, so theoretically AC frequencies lower than 4 kHz may be used for dZ recordings. However, dispersion in myelinated fibres is weak (~30% reduction over a meter nerve) [58]; therefore this effect is not expected to strongly influence the results obtained in simulations.

Impedance changes obtained in the studies on unmyelinated nerves [5], [14], [15], [52], [57] were different from the simulations: significantly lower AC frequencies were needed to record reliable dZ whose amplitudes monotonically decreased with frequency. This is partly because the characteristic frequency of AP in unmyelinated fibres is significantly larger than in the myelinated ones allowing it to be processed using slower carriers (Methods *D*). Also, constant bandwidths were used in the experiments that also affected the dZ-frequency dependence.

### b. Does it offer an explanation on the origin of the dZ?

By analysing the flow of the external injected current as well as the membrane currents and conductances during negative and positive phases of the dZs (observed at 50.7 ms and 51.5 ms, Fig. 6a, Fig. 9a), the model provided insight on their biophysical origin (Fig. 11). In line with expectations, the dZ decrease was due to the opening of ion channels and the significant increase in conductance (Fig. 11a). Conversely, an increase in the dZ was shown to be associated with the change in the flow of the injected current around the recording electrode. This change was due to dynamics of ion channels' opening along the fibre during different phases of AP inducing the current in the external space near the recording electrode to change the direction according to the path of least resistance (Fig. 11e, f). Therefore, the dZ could not be obtained by demodulation of the recorded voltage as it started to depend on the changed current. This conclusion was also supported by measuring the change in the injected current flow through the nodes: it increased during the negative dZ phase but decreased during the positive dZ phase (Fig. 11d).

Thus, the recorded apparent dZ increase did not reflect a real increase in impedance as it was contaminated by the change in the current. However, this apparent increase was reproducible and so may be used in EIT imaging.

   *c. Does it provide results different from the recent model of unmyelinated fibres* [7] *and the passive model* [10]*?*

The dZ in the model of unmyelinated fibres (FCCM) differed from the ones simulated in the current study in two ways. First, the current model showed that dZ could not be measured in a myelinated fibre using any AC frequency. As the dZ frequency band of a modelled myelinated fibre covered a 1–8 kHz range (Fig. 7), high AC measuring frequencies of >4 kHz were necessary for dZ extraction via demodulation (Methods *D*). Second, apart from being the largest at DC, the simulated dZ did not monotonically decrease with frequency. This was because the optimal bandwidths used during band-pass filtering of the recorded signal were determined at each carrier frequency (Fig. 3, Fig. 8) that allowed finding the optimal frequency at which the largest possible dZ could be extracted. On the contrary, these models agreed on the dZ dependence on the sizes and locations of the used electrodes as well as on the presence of the apparent positive dZ changes.

The findings simulated with the passive model [10] did also not match the current one: it included different dependence of the dZ on frequency as well as inability of this model to predict finer properties such as the dZ increase.

3) What recommendations can be given for optimization of imaging myelinated fibres using fast neural EIT?

The main recommendation is to use higher bandwidths for higher carrier frequencies, which have not been previously utilised experimentally. However, since the pure dZ signal decreases with frequency, an optimal combination of AC frequency and bandwidth must be found for each particular nerve; in this study, it was 8 kHz AC with 7.5 kHz bandwidth when subtraction of in- and antiphase signals was performed (Table 3) and this closely agreed with the existing experimental data [4]. Also, the same as for the recent FFCM model, the smallest possible electrodes should be used and the distance between them should be minimized.

*4.3. Limitations and technical difficulties*

The main technical problem, as in the case of the FCCM model, was high computational requirements and therefore long simulation times. Even the use of 2D axisymmetric simplification could not significantly improve the computing speed, which was around 1-2 hours for a single simulation; more than a hundred of those were necessary to test the model with different parameters and obtain necessary statistics. The reason for such low-speed simulations was that the double-cable representation of the fibre was strongly heterogenous which required computation of multiple partial differential equations (2-14 in the appendix) and fulfilling boundary conditions at every connecting point.

Also, a modelling related noise with characteristic frequency of ~18 kHz was found which introduced errors in high-frequency "single-shot" simulations (Fig. 3b). However, this noise was the same at different phases of the injected current and so clearly eliminated in phase-antiphase subtraction case (Fig. 3a).

The FEM approach operates with continuous solution approximation along the fibre. This results in poor convergence at the boundary points connecting nodes with internodes due to sharp transition in potentials in those points (Fig. 4e). To address this issue, the non-uniform mesh was created, where element size decreases towards the boundary (Table 2). Such a mesh included very small elements in the nodal and adjacent regions and was repeated in the external domain for coupling (Fig. 2b) which also significantly increased the time of computation.

*4.4. Work in progress*

Thus, for creating a 3D model of multiple myelinated fibres coupled with the external space, optimisation is critical. One of the approaches currently under development will include multiple 3D single fibre simulations uniformly distributed in the cylinder; one example of such model was accomplished in this study where the fibre was located in the centre of the cylinder (Fig. 2). After the single-fibre simulations are ready, the response of a model with any number of fibres at any position can be reconstructed by interpolation. Parallelizing the simulations using the computer cluster and/or GPUs may be another way to accelerate the development and testing of future, more complex models.

**5. Conclusion**

An accurate FEM model of a myelinated fibre bi-directionally coupled with external space was developed using the experimentally validated space-clamped model of a human sensory fibre. The model allowed simulation of dZs during propagation of an AP and this facilitated determining the optimal parameters for imaging myelinated fibres with EIT.

The dZ decrease simulated with the model was in agreement with experimental data. The model also predicted a small apparent dZ increase and was able to give an explanation to its biophysical origin. The performed simulations allowed finding the optimal bandwidths at each AC frequency for enhancing the efficiency of dZ measurements. Also, optimal currents, as well as sizes and locations of the used electrodes, were determined.

This work also provides the means for optimisation of frequency division multiplexing [59], [60] allowing for simultaneous application of AC at different frequencies which may be used with the predicted optimal parameters to have a possibility of obtaining high quality EIT images of fast neural activity in brain and nerves in real time. The constructed model, together with the previous model of unmyelinated fibres, will serve as a basis for the development of a full model of the nerve which can be used for a range of various applications. They may include the study of biophysical mechanisms of membrane behaviour under a wide range of external conditions, in addition to optimization of various nerve-related technologies such as neural prosthetics or fast neural EIT.

## 6. Acknowledgements

This research was supported by GSK/Verily (Galvani Bioelectronics)-UCL collaboration grant "Imaging and selective stimulation of autonomic nerve traffic using Electrical Impedance Tomography and a non-penetrating nerve cuff".

## Appendix

### 1. Double cable FEM model of a myelinated mammalian sensory fibre

The equations used for simulation of various segments of the fibre were directly derived from the circuit (Fig. 1 in the main text) using Ohm's and Kirchhoff's laws. At the nodes they were:

$$\frac{r_n}{2\rho_i}\left(\frac{\partial^2 V_{ax}}{\partial x^2}+\frac{\partial^2 V_m}{\partial x^2}\right)=C_n\frac{dV_{ax}}{dt}+\sum_{node}I_{ion}(V_{ax}); \quad (2)$$

$$\sum_{node}I_{ion}(V_{ax})=I_{Na}+I_{K_{sn}}+I_{K_{fn}}+I_{Lk_n}+I_{P_n}$$

$$V_m = 0 \quad (3)$$

where $V_{ax}$ and $V_m$ are transmembrane and transmyelin potentials, [$mV$]; $C_n$ is nodal capacitance, [$\mu F/cm^2$]; $r_n$ is the radius of the axon at the node, [$cm$]; $\rho_i$ is the resistivity of the axoplasm, [$kOhm\cdot cm$]; $I_{ion}(V_{ax})$ – ionic currents at the nodal area [$\mu A/cm^2$], which are represented by ion channels from [23]: combined transient and persistent sodium $Na$, fast and slow potassium $K_{sn}$ and $K_{fn}$, Leakage $Lk_n$ and pump $P_n$. Due to absence of myelin sheath at the node, transmyelin potential there equals zero.

At the myelin attachment segments (MYSA) and the internodes (STIN):

$$\frac{r_{ax}}{2\rho_i}\left(\frac{\partial^2 V_{ax}}{\partial x^2}+\frac{\partial^2 V_m}{\partial x^2}\right)=C_{ax}\frac{dV_{ax}}{dt}+\sum_{MYSA/STIN}I_{ion}(V_{ax}); \quad (4)$$

$$\sum_{MYSA/STIN}I_{ion}(V_{ax})=I_{K_{si}}+I_{K_{fi}}+I_h+I_{Lk_i}+I_{P_i}$$

$$\frac{1}{\rho_i}\frac{S_{ax}}{L_m}\left(\frac{\partial^2 V_{ax}}{\partial x^2}+\frac{\partial^2 V_m}{\partial x^2}\right)+\frac{1}{\rho_{pax}}\frac{S_{pax}}{L_m}\frac{\partial^2 V_m}{\partial x^2}=C_m\frac{dV_m}{dt}+\frac{V_m}{\rho_m} \quad (5)$$

where $C_{ax}$ and $C_m$ are capacitances of the axon and myelin sheath, [$\mu F/cm^2$]; $\rho_{pax}$ is the resistivity of the periaxonal space, [$kOhm\cdot cm$]; $r_{ax}$ is the radius of the axon, [$cm$]; $S_{ax}=\pi r_{ax}^2$ and $S_{pax}=\pi(r_{ax}+h)^2-\pi r_{ax}^3$ are the cross-sectional areas of the axon and peri-axonal space at MYSA and STIN segments in [$cm^2$], where $h$ is the width of the periaxonal space at these regions [$cm$]; $L_m$ is the full circumference length of the myelin sheath, [$cm$], its calculation can be found below; $\rho_i$ is the resistivity of the axoplasm, [$kOhm\cdot cm$]; $\rho_m$ is the resistivity of the myelin, [$kOhm\cdot cm^2$]; $I_{ion}(V_{ax})$ – internodal ionic currents, [$\mu A/cm^2$] [23]: slow and fast potassium $K_{si}$ and $K_{fi}$, hyperpolarization activated $h$-channel, leakage $Lk_i$ and pump $P_i$.

Full circumferential length of the myelin sheath was found as the sum of the monotonically rising N = 150 circles of myelin lamellae (Table 2):

$$L_m = 2\cdot 2\pi\sum_{i=1}^{150}r_i = 2\pi\cdot 150\left(r_{ax}+r_{full}+h\right), because$$

$$\sum_{i=1}^{150}r_i = \begin{pmatrix}(r_{ax}+h)+(r_{ax}+h+h_m)+...\\+(r_{ax}+h+(150-1)\cdot h_m)\end{pmatrix}=150\left((r_{ax}+h)+\frac{149}{2}h_m\right) \quad (6)$$

where factor 2 was added as the fibre contained 2 membranes per lamella. The radii of the circles rose from $r_1 = r_{ax}+h$ to $r_{150} = d_{full}/2 = 8\ \mu m$; $h_m$ is a distance between lamellae: $h_m = (r_{150}-r_1)/150$ (Table 3).

Ionic currents at the node were as follows:

$$I_{Na}=\frac{P_{Na}\frac{VF^2}{RT}(m^3h+P_{NaP}m_p^3)\left(S_{Na}\left([Na]_0-[Na]_i e^{\frac{VF}{RT}}\right)+(1-S_{Na})\left([K]_0-[K]_i e^{\frac{VF}{RT}}\right)\right)}{1-e^{\frac{VF}{RT}}} \quad (7)$$

$$I_{K_{sn}}=\bar{g}_{K_{sn}}s\left(V-V_{K_s}\right)$$

$$I_{K_{fn}}=\bar{g}_{K_{fn}}n^4\left(V-V_{K_f}\right)$$

$$I_{Lk_n}=\bar{g}_{Lk_n}\left(V-V_{rest,n}\right)$$

At the internode:

$$I_{K_{si}}=\bar{g}_{K_{si}}s_i\left(V-V_{K_s}\right)$$

$$I_{K_{fi}}=\bar{g}_{K_{fi}}n_i^4\left(V-V_{K_f}\right) \quad (8)$$

$$I_h=\bar{g}_h q\left(V-V_h\right)$$

$$I_{Lk_i}=\bar{g}_{Lk_i}\left(V-V_{rest,i}\right)$$

Reversal potentials were calculated as:

$$V_x=\frac{\ln\left(\frac{[K]_0+S_x[Na]_0-S_x[K]_0}{[K]_i+S_x[Na]_i-S_x[K]_i}\right)}{F/RT}, x=Na,K_s,K_f,h \quad (9)$$

In the equations, $V\equiv V_{ax}$ from the previous equations, [$mV$]; $F$ – Faraday's constant, [$\mu A\cdot ms/mM$], $P_{Na}$ – permeability of $Na$ channels, [$cm^3/ms$]; $P_{NaP}$ – percent of persistent $Na$ channels; $[Na]_{0,i}$ and $[K]_{0,i}$ are sodium and potassium concentrations outside and inside the axon, [$mM$]; $R$ – gas constant, [$pJ/mM\cdot K$]; T – temperature, [$K$]; $S_x$ – selectivity of $Na$, $K$, $h$ channels; $\bar{g}_{K_{sn,i}},\bar{g}_{K_{fn,i}},\bar{g}_{Lk_{n,i}},\bar{g}_h$ – maximal conductances of potassium slow and fast channels, leakage and h-channel at the node and internode, [$1/(k\Omega\cdot cm^2)$]; $V_{K_s},V_{K_f},V_h$ –reversal potentials of corresponding ion channels explained by (7), [$mV$]; $V_{rest,\ n/i}$ – nodal and internodal resting potentials, [*mV*]; *m, m_p, h, s, n, s_i, n_i, q* – gating variables explained by corresponding gating equations of the form (9) with the coefficients $\alpha_x = \alpha_x(V_{ax})$ and $\beta_x = \beta_x(V_{ax})$ given in [23].

$$\frac{dm}{dt} = \alpha_m(V_{ax}) \cdot (1-m) - \beta_m(V_{ax}) \cdot m \qquad (10)$$

and similarly, for all other gating variables. Initial values were found by solving (9) for *m* (and other gating variables) when $dm/dt = 0$ which designates stabilization of a gating variable in time.

All the electrical parameters of the modelled fibre included in the equations above are presented in the Table 4 below.

TABLE 4
ELECTRICAL PARAMETERS OF THE MODEL

| Parameter | Value |
|---|---|
| $C_n$ | 1 µF/cm² |
| $C_{ax}$ | 1 µF/cm² |
| $C_m$ | 1.9 pF/cm² |
| $\rho_i$ | 70 Ω·cm |
| $\rho_{pax}$ | 70 Ω·cm |
| $\rho_m$ | 2.8 MΩ·cm² |
| $[K]_0$ | 4.5 mM |
| $[K]_i$ | 155 mM |
| $[Na]_0$ | 144.2 mM |
| $[Na]_i$ | 9 mM |
| $V_h$ | -54.85 mV |
| $V_{Ks,f}$ | -94.5 mV |
| $V_{Na}$ | 45.6 mV |
| $V_{rest,n}$ | -80.3 mV |
| $V_{rest,i}$ | -81.3 mV |
| $T$ | 310 K |
| $P_{Na}$ | 2.27·10⁻⁵ cm³/ms |
| $P_{NaP}$ | 1.07 |
| $S_{Na}$ | 0.9 |
| $S_K$ | 0 |
| $S_h$ | 0.097 |
| $\bar{g}_{K_{sn}}$ | 109.47 mS/cm² |
| $\bar{g}_{K_{fn}}$ | 87.58 mS/cm² |
| $\bar{g}_{Lk_n}$ | 9.78 mS/cm² |
| $I_{P_n}$ | -170.64 µA/cm² |
| $\bar{g}_{K_{si}}$ | 5.81·10⁻³ mS/cm² |
| $\bar{g}_{K_{fi}}$ | 0.69 mS/cm² |
| $\bar{g}_h$ | 1.37·10⁻² mS/cm² |
| $\bar{g}_{Lk_i}$ | 1.22·10⁻² mS/cm² |
| $I_{P_i}$ | -0.018 µA/cm² |
| $S_{ax}$ | 1.27·10⁻⁶ cm² |
| $S_{pax,\,stin}$ | 7.98·10⁻¹¹ cm² |
| $S_{pax,\,mysa}$ | 7.98·10⁻¹² cm² |
| $L_m$ | 1.35 cm |

*2. FEM model coupled with external space*

Equations describing the bi-directionally coupled FEM model of the myelinated fibre are presented below.

First, Laplace's equation (10) represents volume conduction in the extracellular space where $\sigma_e$ is its conductivity, $[1/(kOhm \cdot cm)]$ and $V_e$ is the potential, $[mV]$.

$$-\nabla(\sigma_e(\nabla V_e)) = 0, \text{ in V} \qquad (11)$$

The second set of equations describes how the main equations for the fibre (1), (3) and (4) are transformed by addition of external stimulation for each compartment represented by the activating function $d^2V_e(x,t)/dx^2$; these equations can also be obtained by applying Kirchhoff's laws to the circuit (Fig.1 in the main text).

$$\begin{aligned}
\frac{r_n}{2\rho_i}\left(\frac{\partial^2 V_{ax}}{\partial x^2} + \frac{\partial^2 V_m}{\partial x^2} + \frac{\partial^2 V_e}{\partial x^2}\right) &= C_n \frac{dV_{ax}}{dt} + \sum_{node} I_{ion}(V_{ax}); \\
\frac{r_{ax}}{2\rho_i}\left(\frac{\partial^2 V_{ax}}{\partial x^2} + \frac{\partial^2 V_m}{\partial x^2} + \frac{\partial^2 V_e}{\partial x^2}\right) &= C_{ax}\frac{dV_{ax}}{dt} + \sum_{MYSA/STIN} I_{ion}(V_{ax}); \\
\frac{1}{\rho_i}\frac{S_{ax}}{L_m}\left(\frac{\partial^2 V_{ax}}{\partial x^2} + \frac{\partial^2 V_m}{\partial x^2} + \frac{\partial^2 V_e}{\partial x^2}\right) &+ \frac{1}{\rho_{pax}}\frac{S_{pax}}{L_m}\left(\frac{\partial^2 V_m}{\partial x^2} + \frac{\partial^2 V_e}{\partial x^2}\right) = \\
&= C_m \frac{dV_m}{dt} + \frac{V_m}{\rho_m}
\end{aligned} \qquad (12)$$

The final equations show the flux of the transmembrane current from the node or myelin (including MYSA and STIN).

$$\begin{aligned}
I_n|_\Gamma &= \sigma_e \nabla V_e \cdot \mathbf{n} = C_n \frac{dV_{ax}}{dt} + \sum_{node} I_{ion}(V_{ax}), \text{ on } \Gamma_{node} \\
I_m|_\Gamma &= \sigma_e \nabla V_e \cdot \mathbf{n} = C_m \frac{dV_m}{dt} + \frac{V_m}{\rho_m}, \text{ on } \Gamma_{myelin}
\end{aligned} \qquad (13)$$

The constant sinusoidal current applied to the external injecting electrodes was as follows:

$$I_{inj} = \pm I_{amp} \cdot \sin(2\pi \cdot f \cdot t + \varphi) \qquad (14)$$

where $I_{amp}$ is an amplitude expressed in the current density terms – $[\mu A/cm^2]$, *f* is its frequency [*kHz*], $\varphi$ – phase [*rad*], *t* – time [*ms*].

The scheme of operation of the developed coupled model repeated the one in FCCM [7] and included two simultaneous simulations – with and without the injected current. Simulation of electric field generated by the fibre when no current was injected ($V_e^0$) was done via additional 1D and 2D-axisymmetric geometries; their difference with the electric field simulated with the injected current ($V_e$) was applied to the nerve fibre ($V_e$-$V_e^0$) at each time step. Finally, the resultant transmembrane current was coupled from the fibre to the main geometry.